\documentclass[11pt]{article}
\usepackage{amsfonts}
\usepackage{pstricks}
\usepackage{pst-node}
\usepackage{epsfig}
\usepackage{epsfig}

\begin{document}

\def\ba{\begin{eqnarray}}
\def\ea{\end{eqnarray}}
\def\w{\wedge}
\def\d{\mbox{d}}
\def\D{\mbox{D}}

\begin{titlepage}
\title{\bf Gauge Approach to The Symmetric Teleparallel Gravity}
\author{ Muzaffer Adak  \\
 {\small Department of Physics, Faculty of Arts and Sciences,} \\
 {\small Pamukkale University, 20017 Denizli, Turkey} \\
 {\small \tt madak@pau.edu.tr}
 }
  \vskip 1cm
\date{28 August 2018}
\maketitle

 \thispagestyle{empty}

\begin{abstract}
 \noindent
We discuss a gauge invariant gravity model in a non-Riemannian geometry in which
the curvature and the torsion both are zero, the nonmetricity is nonzero. We also argue
that only a metric ansatz is enough to start finding solutions to
the field equations. As an application we obtain explicitly a conformally flat solution.

\vskip 1cm


% 11.15.Kc    (Gauge field theory) Classical and semiclassical techniques
% 04.50.Kd    Modified theories of gravity
% 02.40.Yy    Geometric mechanics
\end{abstract}
\end{titlepage}

\section{Introduction}

The gauge theory is a successful approach in the modern physics for
explaining the nature. For example, the electromagnetic and the weak
nuclear interactions have been combined through $SU(2) \otimes
U(1)$-gauge theory; the electroweak theory. $SU(3) \otimes SU(2)
\otimes U(1)$-gauge theory which is one step advance of the same
approach contains strong nuclear interactions as well; the so-called standard
model \cite{griffiths1987}.

The gauge theory may be summarized briefly as follows. A physical entity  is
represented by a field. A Lagrangian for the field containing kinetic and
mass terms is written. The transformation rules of the field is defined
and the Lagrangian is required to be invariant first globally and
then locally under that transformation. The local invariance is
usually broken because of the derivative in the kinetic term. To
restore the invariance, a new field (gauge potential) and its transformation
rules are introduced. Since this gauge potential is required to be dynamic, the kinetic and mass terms of the gauge potential
are added to the Lagrangian. Then the local invariance of the new
Lagrangian under the defined transformations is checked. If the
local invariance is lost either the term breaking invariance is
dismissed or a new field enters the scheme. Finally, the group
structure of the set of transformation elements is investigated, see the second column of the Table \ref{table:1}.

In the literature, there are different gauge theoretical approaches to
describe gravitation. Firstly the General Relativity (GR) was written in
pseudo-Riemannian geometry in which the field representing the gravitation is the  metric (or correspondingly curvature), the field
representing the gauge potential is the connection and the gauge group is the Lorentz group. Meanwhile, there are strong reasons motivated by recent astrophysical and cosmological observations that the Einstein's general relativity theory needs to be modified. In the
second group, there are gravity models containing the square of
the curvature in order to obtain the Yang-Mills-type terms. Here again the
geometry is pseudo-Riemannian, the metric/curvature represents the gravity and
the connection represents the gauge field. This time, however, the fourth
order non-linear differential equations of the metric functions appear
\cite{stephenson1958}-\cite{benn1981}. This result
gives rise to some troubles in quantization. In the third group, geometry is enlarged to the non-Riemannian geometries
\cite{dereli1994}-\cite{adak2007}.

In general a geometry is classified in accordance with the three tensors, the so-called the curvature, the torsion and the nonmetricity. If all three are zero, it is the Minkowskian geometry. If only the curvature is nonzero, the other two are vanishing, it is Riemannian geometry. If only the nonmetricity is zero, the other two are nonvanishing, it is called the Riemann-Cartan geometry. If only the torsion is nozero, other two are vanishing, it is the Weitzenb\"{o}ck (or teleparallel) geometry. If only the nonmetricity is nonzero, the others are zero, it is called symmetric teleparallel geometry. If all three are nonzero, there is not a consensus on the nomenclature; the non-Riemannian geometry \cite{benn1981} or the metric affine geometry \cite{hehl1995}.

In this work, we propose a gravity model in a geometry in which
only the nonmetricity is nonzero, namely the Symmetric Teleparallel Gravity
(STPG) \cite{nester1999}-\cite{adak2013}. Here the interpretation is
then clear. The gravitational and the gauge fields are represented by the metric
and the connection, respectively. Another important contribution of this work to the STPG literature, in which a metric ansatz and
a connection ansatz are done independently, is to show that
only a metric ansatz is enough to begin to look for solutions. Details can be found in Section \ref{kisimSTPG}.
Finally, after discussing some kinetic and mass terms for the gauge
field, we find a conformally flat solution to the field
equations derived from a gauge invariant STPG Lagrangian via
the independent variations.

\section{ The Mathematical Preliminaries}

The spacetime, in general, is denoted by the triple $\{ M,g,\nabla\}$ where $M$
is the four-dimensional orientable and differentiable manifold, $g$ is the (0,2)-type
symmetric metric tensor, $\nabla$ is the connection. Let $
\{x^\alpha(p)\}, \ \alpha =\hat{0},\hat{1},\hat{2},\hat{3},$ be
the coordinate functions of the coordinate system at any point $p \in
M$. This coordinate system forms the reference frame denoted by $
\{\frac{\partial}{\partial x^\alpha}(p)\}$ or shortly
$\partial_\alpha (p) $, the so-called {\it coordinate frame}. This
frame is a set of basis vectors at point $p$ for the tangent space
$T_p(M)$. Similarly, the set of differentials of coordinate
functions $\{ \d x^\alpha (p) \}$ forms the {\it coordinate
co-frame} at the point $p$ for the cotangent space $T_p^*(M)$, the
so-called base co-vectors. Interior product of $\partial_\alpha$
and $dx^\alpha$ is given by Kronecker symbol
 \begin{eqnarray}
          \d x^\alpha \left( \frac{\partial}{\partial x^\beta} \right) \equiv
           \imath_{\beta} \d x^\alpha =
           \delta^\alpha_\beta \; . \label{2.1}
 \end{eqnarray}
Any linearly independent vectors can be made orthonormal. Let $
\{X_a\},\ a=0,1,2,3,$ be the orthonormal set of the vectors, the so-called
{\it orthonormal frame}. In this case, the metric satisfies
$g(X_a,X_b)=\eta_{ab}$ where $\eta_{ab}=\mbox{diag}(-1,1,1,1)$.
Let $\{e^a\}$ be the dual set of $\{X_a\}$, the so-called {\it
orthonormal co-frame}
 \begin{eqnarray}
          e^a(X_b)\equiv \imath_be^a=\delta^a_b \; . \label{2.2}
 \end{eqnarray}
This is another manifestation of (\ref{2.1}). On the other hand, it is always possible to work in a {\it mixed frame} in between the coordinate frame, $\d e^\alpha = \d^2 x^\alpha =0$, and the orthonormal frame, $\d \eta_{ab}=0$, in which the exterior derivative of the co-frame and the components of metric both are nonzero, $\d e^\alpha \neq 0$ and $\d g_{\alpha \beta} \neq 0$.

Unless expressly stated, in this work we adhere the notation. Greek indices are called the {\it
coordinate indices}; $\alpha, \beta, \ldots = \hat{0}, \hat{1},
\hat{2}, \hat{3}$. The Latin ones are called the {\it orthonormal
indices}; $a,b,\ldots = 0,1,2,3$. One can pass from the coordinate
frame to the orthonormal frame and the vice versa via the vierbein
${h^{\alpha}}_a(p)$
 \begin{eqnarray}
  X_a(p) = {h^{\alpha}}_a(p)\partial_{\alpha}(p) \; .
 \end{eqnarray}
In order $\{ X_a \}$ to form a basis, the vierbein must be
nondegenerate; i.e. $\mbox{det}{h^{\alpha}}_a(p) \neq 0$. Similarly,
 \begin{eqnarray}
  e^a(p) = {h^a}_\alpha (p) \d x^{\alpha}(p) \; .
 \end{eqnarray}
Besides, because of (\ref{2.1}) and (\ref{2.2}) it is written
 \begin{eqnarray}
 {h^{a}}_\alpha (p){h^\beta}_{a}(p) = \delta^\beta_\alpha \; , \quad  {h^{\alpha}}_a(p){h^b}_{\alpha}(p) =
 \delta^b_a \; .
 \end{eqnarray}

The connection $\nabla$ is determined by the connection 1-forms
${\Lambda^a}_b$. The orientation of the manifold is fixed by the Hodge map
$*1=e^0 \wedge e^1 \wedge e^2 \wedge e^3 $ where $\wedge$ denotes the
exterior product. From now on we make use of the abbreviation $e^{ab
\cdots } \equiv e^a \wedge e^b \wedge \cdots $. The Cartan structure
equations define the nonmetricity 1-forms, the torsion 2-forms and
the curvature 2-forms. They are written explicitly in a {\it mixed frame} respectively as follows
 \ba
     Q_{\alpha \beta} &:=& - \frac{1}{2} \D g_{\alpha \beta}
                      = \frac{1}{2} (-\d g_{\alpha \beta} + \Lambda_{\alpha \beta}+\Lambda_{ \beta \alpha}) \, , \label{nonmet}\\
     T^\alpha &:=& \D e^\alpha = \d e^\alpha + {\Lambda^\alpha}_\beta \wedge e^\beta \, , \label{tors}\\
     {R^\alpha}_\beta &:=& \D {\Lambda^\alpha}_\beta := \d {\Lambda^\alpha}_\beta
                  + {\Lambda^\alpha}_\gamma \wedge {\Lambda^\gamma}_\beta \label{curva}
 \ea
where $ \d $ and $ \D$ are the exterior derivative and the covariant
exterior derivative, respectively.   They satisfy the Bianchi
identities:
 \ba
       \D Q_{\alpha \beta} &=& \frac{1}{2} ( R_{\alpha \beta} +R_{\beta \alpha}) \, , \label{bianc:0} \\
       \D T^\alpha    &=& {R^\alpha}_\beta \wedge e^\beta \, , \label{bianc:1} \\
       \D {R^\alpha}_\beta &=& 0  \, . \label{bianc:2}
 \ea

  \subsection{The Decomposition of the Full Connection } \label{sec:dec-con}

In a {\it mixed frame} the full connection 1-forms can be decomposed uniquely as follows
\cite{hehl1995},\cite{tucker1995}:
 \ba
     {\Lambda^\alpha}_\beta =  \underbrace{(g^{\alpha \gamma}\d g_{\gamma_\beta} + {p^\alpha}_\beta)/2 + {\omega^\alpha}_\beta}_{Metric}
     + \underbrace{{K^\alpha}_\beta}_{Torsion} + \underbrace{{q^\alpha}_\beta  + {Q^\alpha}_\beta}_{Nonmetricity}   \label{connect:dec}
 \ea
where ${\omega^\alpha}_\beta$ the Levi-Civita connection 1-forms
 \ba
     {\omega^\alpha}_\beta \wedge e^\beta = -\d e^\alpha  \, , \label{LevCiv}
 \ea
$K^\alpha{}_\beta$ the contortion tensor 1-forms,
 \ba
   {K^\alpha}_\beta  \wedge e^\beta = T^\alpha  \, , \label{contort}
 \ea
$p_{\alpha \beta}$ and $q_{\alpha \beta}$ the anti-symmetric 1-forms
 \ba
  & & p_{\alpha \beta} = -( \imath_\alpha  \d g_{\beta \gamma } ) e^\gamma + ( \imath_\beta \d g_{\alpha \gamma})
    e^\gamma \, . \label{p:ab} \\
   & & q_{\alpha \beta} = -( \imath_\alpha  Q_{\beta \gamma } ) e^\gamma + ( \imath_\beta Q_{\alpha \gamma})
    e^\gamma  \, , \label{q:ab}
 \ea
This decomposition is self-consistent. To see that it is enough to
multiply (\ref{connect:dec}) from right by $e^\beta$ and to use
definitions above. While moving indices vertically in front of
both $\d$ and $\D$, special attention is needed because $\d
g_{\alpha \beta} \neq 0$ and $\D g_{\alpha \beta} \neq 0$.
The symmetric part of the full connection comes from (\ref{nonmet})
 \ba
  \Lambda_{(\alpha \beta)} = Q_{\alpha \beta } + \frac{1}{2} \d g_{\alpha \beta } \label{connect:sym}
 \ea
and the remainder is the anti-symmetric part
 \ba
  \Lambda_{[\alpha \beta]} = \frac{1}{2} p_{\alpha \beta} + \omega_{\alpha \beta} + K_{\alpha \beta} + q_{\alpha \beta}  \, . \label{connect:ansym}
 \ea

If only $Q_{\alpha \beta}=0$, the connection is metric compatible. If
both $Q_{\alpha \beta}=0$ and $T^\alpha =0$, the connection is
the Levi-Civita. If $\d g_{\alpha \beta }=0$, we denote the metric
components as $\eta_{ab}$ and call it as the {\it orthonormal metric}.
In this case,  the decomposition of the full connection takes the form
 \ba
 \Lambda_{ab} = \omega_{ab} + K_{ab} + q_{ab}
          +  Q_{ab} \, . \label{connect:on}
 \ea
In the literature there are works preferring orthonormal frames
\cite{dereli1994}, coordinate frames \cite{hehl1995} and neither
of the two \cite{adak2007}. In calculations the following
identities are useful:
 \ba
    \D*e_\alpha &=& - Q \wedge *e_\alpha + * e_{\alpha \beta} \wedge T^\beta \; , \label{ident:1}\\
    \D*e_{\alpha \beta}  &=& -  Q \wedge *e_{\alpha \beta} +  * e_{\alpha \beta \gamma} \wedge T^\gamma \; , \label{ident:2} \\
   \D*e_{\alpha \beta \gamma} &=& -  Q \wedge *e_{\alpha \beta
   \gamma} +  * e_{\alpha \beta \gamma \sigma} \wedge T^\sigma  \; , \label{ident:3} \\
    \D*e_{\alpha \beta \gamma \sigma} &=& -  Q \wedge *e_{\alpha \beta \gamma \sigma}  \label{ident:4}
 \ea
where $ Q := Q^\alpha{}_\alpha = g^{\alpha \beta} Q_{\alpha
\beta}$ is the trace 1-form of nonmetricity.

 \section{The Symmetric Teleparallel Gravity} \label{kisimSTPG}

In the orthonormal frame the Riemannian geometry is defined by the configuration
 \ba
  & &  Q_{ab} = \frac{1}{2} (\Lambda_{ab} + \Lambda_{ba})=0 \, , \nonumber \\
  & &   T^a = \d e^a +{\Lambda^a}_b \wedge e^b = 0 \, , \nonumber \\
   & &   {R^a}_b = \d {\Lambda^a}_b + {\Lambda^a}_c \wedge {\Lambda^c}_b \neq 0 \, .
 \ea
The first two equations yield a set of linear equations for the connection 1-forms and they can be solved uniquely in terms of the orthonormal co-frame which are obtained from the metric as
 \ba
     \Lambda_{ab} = -\Lambda_{ba} = \frac{1}{2} [-\iota_a \d e_b + \iota_b \d e_a + (\iota_a\iota_b \d e_c) e^c] \, .
 \ea
This is called the Levi-Civita connection 1-forms and denoted by $\omega_{ab}$ in this work. On the other hand, in the STPG models only the nonmetricity is nonzero
  \ba
  & &  Q_{ab} = \frac{1}{2} (\Lambda_{ab} + \Lambda_{ba}) \neq 0 \, , \nonumber \\
   & &  T^a = \d e^a +{\Lambda^a}_b \wedge e^b = 0 \, , \nonumber \\
   & &   {R^a}_b = \d {\Lambda^a}_b + {\Lambda^a}_c \wedge {\Lambda^c}_b = 0 \, . \label{STPG1}
 \ea
Here as the second equation gives a set of linear equations for $\Lambda_{ab}$, the third one reads a set of nonlinear partial differential equations for them. Therefore one can not obtain a unique solution in terms of the orthonormal co-frame, i.e. the metric, for the full connection 1-forms like that done in the Riemannian geometry. Thus we state a proposition here.

{\it Proposition:} The conditions (\ref{STPG1})
can be satisfied by starting in the coordinate frame and by suitably choosing the full connection. Let us prove this statement step by step:
\begin{itemize}
    \item \verb"Step 1": Write the metric $ g= g_{\alpha \beta}(x) \d x^\alpha
           \otimes \d x^\beta $ where $x$ denotes the coordinate functions.
    \item \verb"Step 2": Fix the gauge, that is, choose the coordinate co-frame $ e^\alpha = \d x^\alpha $ and the full connection
           1-forms as  $
           \Lambda^\alpha{}_\beta =0 $. So, $ R^\alpha{}_\beta =0 $, $ T^\alpha =0$, $ Q_{\alpha \beta} = - \frac{1}{2} \d
           g_{\alpha \beta}  \neq 0 $ because of the equations (\ref{nonmet})-(\ref{curva}).
    \item \verb"Step 3": Determine the vierbien and its inverse by using $e^a (x) =
           h^a{}_\alpha (x) \d x^\alpha $.
    \item \verb"Step 4": In the orthonormal frame,
           calculate the full connection $\Lambda^a{}_b = h^a{}_\alpha
          \Lambda^\alpha{}_\beta h^\beta{}_b + h^a{}_\alpha \d h^\alpha{}_b
          = h^a{}_\alpha \d h^\alpha{}_b \neq 0 $, the curvature $ R^a{}_b =
           h^a{}_\alpha R^\alpha{}_\beta h^\beta{}_b = 0$, the torsion $ T^a =
           h^a{}_\alpha T^\alpha =0 $ and the nonmetricity  $Q^a{}_b = h^a{}_\alpha
            Q^\alpha{}_\beta h^\beta{}_b  \neq 0 $. This result corresponds to
            $\omega_{ab} + q_{ab} =0$ in the decomposition (\ref{connect:on})
            which means $\Lambda_{ab} = Q_{ab}$ together with $K_{ab}=0$.
\end{itemize}

 \subsection{The Gauge Approach}

In the symmetric teleparallel geometry the identities (\ref{bianc:0})-(\ref{bianc:2})
yield only one nontrivial identity,
 \ba
   \D Q_{ab} =0 \; .
 \ea
In the gauge approach to the STPG we firstly propose the Lagrangian 4-form
 \ba
     \mathcal{L}' = -\frac{\kappa}{8} \d g_{\alpha \beta} \wedge *\d g^{\alpha \beta}
     + M *1 \label{LagranjGlobal}
 \ea
where $\kappa$ and $M$ are constants. Here let $g_{\alpha \beta}$
represent the gravitational field. Meanwhile, the mass term would be
written as $M *1 = \frac{M}{4} g_{\alpha \beta} \wedge *g^{\alpha
\beta} $. This Lagrangian is invariant under the {\it global gauge
transformation}
 \ba
    g_{\alpha \beta} = g_{\alpha' \beta'}
    L^{\beta'}{}_\beta L^{\alpha'}{}_\alpha \quad , \quad
   g^{\alpha \beta} = L^{\beta}{}_{\beta'} L^{\alpha}{}_{\alpha'}
     g^{\alpha' \beta'}  \label{DonusumMetrik}
 \ea
in the sense that the transformation elements are independent of the coordinates, $\d
L^{\beta'}{}_\beta =0 $. Here $g_{\alpha \beta} g^{\alpha \gamma}
= \delta^\gamma_\beta$  causes the conditions
$L^{\beta}{}_{\beta'} L^{\beta'}{}_{\alpha} = \delta_\alpha^\beta$
and $L^{\beta'}{}_{\beta} L^{\beta}{}_{\alpha'} =
\delta_{\alpha'}^{\beta'}$ on the transformation elements. Now in
accordance  with the gauge theory, let us impose the condition of the  local invariance to the Lagrangian 4-form: $ L^{\beta'}{}_\beta =
L(x)^{\beta'}{}_\beta$. In this case $\mathcal{L}'$ is not
invariant because of the terms $\d L^{\beta'}{}_{\beta} (x) \neq
0$. Thus, in order to restore the invariance we are obligated to add terms such as $\d
g_{\alpha \beta} \wedge *\Lambda^{\alpha \beta}$ and
$\Lambda_{\alpha \beta} \wedge *\Lambda^{\alpha \beta}$ etc., such
that
 \ba
     \mathcal{L} = \frac{\kappa}{2} Q_{\alpha \beta} \wedge *Q^{\alpha \beta}
     + M *1 \label{LagranjYerel}
 \ea
where $Q_{\alpha \beta} := -\frac{1}{2} \D g_{\alpha \beta}$ and
$Q^{\alpha \beta} := \frac{1}{2} \D g^{\alpha \beta}$. This new
Lagrangian is invariant under the {\it local gauge transformations}
 \ba
   \Lambda^\alpha{}_\beta = L(x)^\alpha{}_{\alpha'} \Lambda^{\alpha'}{}_{\beta'}  L(x)^{\beta'}{}_\beta + L(x)^\alpha{}_{\alpha'}
                               \d L(x)^{\alpha'}{}_\beta \label{DonusumBaglantiYerel}
 \ea
The group formed by the elements $L^\alpha{}_{\alpha'} (x)$ is the general linear group because the full connection 1-form $\Lambda_{ab}$ contains $4^2=16$ 1-forms in four dimensions because of the nonmetricity \cite{benn1982},\cite{cornwell1997}. We compare the STPG with the minimally coupled Dirac-Maxwell theory in the Table \ref{table:1}.

\begin{table}
\caption{Here $\bar{\psi}$ denotes the Dirac conjugate of the Dirac field, $\psi$, the Clifford algebra valued 1-form, $\gamma = \gamma_a e^a$, is defined in terms of the generators, $m$ is the mass of the Dirac particle, $F$ is the Maxwell 2-form, $U(1)$ denotes the unitary group and $GL(4)$ denotes the general linear group.}
\label{table:1}
\begin{tabular}{ | c | c| c | }
\hline
      & Maxwell-Dirac & STPG \\
\hline
\hline
Field & $\psi (x)$ & $g=g_{\alpha \beta}(x) $d$ x^\alpha \otimes $d$ x^\beta $ \\
\hline
Globally gauge & $\frac{i}{2} ( \bar{\psi} *\gamma \wedge $d$\psi + $d$\bar{\psi} \wedge *\gamma \psi ) $ & $\frac{\kappa}{8} $d$g_{\alpha \beta} \wedge *$d$g^{\alpha \beta} $\\
invariant Lagrangian & $+i m \bar{\psi} \psi *1$ & $ + \frac{M}{4} g_{\alpha\beta} g^{\alpha \beta}*1$ \\
\hline
Global gauge & $\psi \rightarrow e^{i\theta} \psi$ where & $g \rightarrow L^T g L$ where \\
transformation &$\theta$ is any real number & $L$ contains real numbers\\
\hline
Locally &$\frac{i}{2}\left( \bar{\psi} *\gamma \wedge \mathcal{D} \psi + \mathcal{D}\bar{\psi} \wedge *\gamma \psi \right) $ & $\frac{\kappa}{8} $D$g_{\alpha \beta} \wedge *$D$g^{\alpha \beta} + M*1 $\\
gauge &$+i m \bar{\psi} \psi *1 + F \wedge *F$ where & $  + {R^\alpha}_\beta \wedge *{R^\beta}_\alpha$  where\\
invariant & $\mathcal{D} \psi := (d-iA)\psi$, & $  $D$ g_{\alpha \beta} := \frac{1}{2} (-$d$ g_{\alpha \beta} + \Lambda_{\alpha \beta}+\Lambda_{\beta \alpha})$,\\
Lagrangian & $F:= $d$A$ & ${R^\alpha}_\beta  := $d$ {\Lambda^\alpha}_\beta
                  + {\Lambda^\alpha}_\gamma \wedge {\Lambda^\gamma}_\beta $\\
\hline
Local gauge &$\psi \rightarrow e^{i\theta(x)} \psi$, & $g \rightarrow L^T(x) g L(x)$, \\
transformations & $A \rightarrow A+$d$\theta$ & $\Lambda \rightarrow L^{-1} \Lambda L + L^{-1}$d$L$\\
\hline
Gauge group & $U(1)$ & $GL(4)$\\
\hline
Gauge potential & $A$ & $\Lambda $\\
\hline
Bianchi & $\mathcal{D}^2\psi =-iF\psi$ &$$D$^2g_{\alpha \beta}=-( R_{\alpha \beta} +R_{\beta \alpha})$ \\
identities &  $$d$F=0$ &  $$D$ {R^\alpha}_\beta=0 $ \\
\hline
\end{tabular}
\end{table}

In the gauge theoretical approach, the kinetic and the mass terms of the gauge
field are expected to be added to $\mathcal{L}$. The first
possibility would be $ \mathcal{L}_1 \sim c_1 \D
\Lambda^\alpha{}_\beta \wedge *\D \Lambda^\beta{}_\alpha + c_2
\Lambda^\alpha{}_\beta \wedge *\Lambda^\beta{}_\alpha$. Here,
however, the first term reminds the expression $R^\alpha{}_\beta
\wedge *R^\beta{}_\alpha$ and it is zero in the STPG. And also since
the second term is not invariant under the transformation
(\ref{DonusumBaglantiYerel}) it must be discarded. Another candidate
would be the Chern-Simons term
 $
     K = {\Lambda^a}_b \wedge \d {\Lambda^b}_a + \frac{2}{3}
     {\Lambda^a}_b \wedge {\Lambda^b}_c \wedge {\Lambda^c}_a
 $
invariant under the transformation (\ref{DonusumBaglantiYerel}).
In this case, the locally gauge invariant Lagrangian in four-dimensions
would be $ \mathcal{L}_2 \sim \d \theta \wedge K + \frac{1}{2}
\d \theta \wedge * \d \theta $. Here again by following the gauge
approach, we add the kinetic term of newly introduced scalar
field $\theta$. The mass term, $\theta^2 *1$, may be put into $M$ term of $\mathcal{L}$. But, $\mathcal{L}_2$ can be rewritten in the form of $ \mathcal{L}_2 \sim \theta {R^a}_b \wedge {R^b}_a + \frac{1}{2} \d \theta \wedge
* \d \theta $ because of the property $\d K = - {R^a}_b \wedge {R^b}_a$
when the exact form is discarded. The first term is again zero in
the STPG and then no reason leaves to keep the second term.

 \subsection{The Field Equations}

We prefer working in the orthonormal gauge in this section. That is
because the orthonormal metric components are made of $0,\pm 1$ and
their variations are all zero. Thus, only the independent variations of $\mathcal{L}$ with respect to the orthonormal co-frame and
the connection are needed.
 \ba
 \mathcal{L} = \frac{\kappa}{2} Q_{ab} \wedge *Q^{ab}
     + M *1 + \lambda_a \wedge T^a + R^a{}_b \wedge \rho_a{}^b \label{LagranjYerel2}
 \ea
where $\lambda_a$ and $\rho_a{}^b$ the Lagrangian multiplier 2-forms
giving constrains, respectively, $T^a=0$ and $R^a{}_b =0$. In
addition to these equations we obtain from the connection variation
 \ba
     \lambda_a \wedge e^b + \D \rho_a{}^b = -\Sigma_a{}^b \; ,
     \label{AlanDenk1}
 \ea
and from the co-frame variation
 \ba
      M*e_a + \D \lambda_a = - \tau_a \label{AlanDenk2}
 \ea
where
 \ba
    \Sigma_a{}^b &=& \frac{\kappa}{2} *(Q_a{}^b + Q^b{}_a) \; , \label{Sigma_a^b}\\
    \tau_a &=& -\frac{\kappa}{2} [(\imath_a Q^{bc}) *Q_{bc} + Q^{bc} \wedge (\imath_a
    *Q_{bc})] \; . \label{tauA}
 \ea
We can surely say that the theory (\ref{LagranjYerel2}) is not viable, because it has a ghost in its spectrum. The Minkowski solution is not stable. This has been well established in several studies during the past year, and one of these computations is reviewed in Section 3 of \cite{koivisto2018}.
In principle, the Lagrangian multipliers are solved from
Eqn(\ref{AlanDenk1}) and the results are substituted into
Eqn(\ref{AlanDenk2}). We notice that what we need in the second
equation is only $\D \lambda_a$ rather than $\lambda_a$ or
$\rho_{ab}$. Thus, by taking the exterior derivative of
Eqn(\ref{AlanDenk1})
 \ba
    \D \lambda_a \wedge e^b = -\D \Sigma_a{}^b \; . \label{AlanDenk3}
 \ea
Here we use $\D e^a =T^a =0$ ve $\D^2 \rho_a{}^b = R^b{}_c \wedge
\rho_a{}^c - R^c{}_a \wedge \rho_c{}^b =0$. Now, after multiplying
Eqn(\ref{AlanDenk2}) by $ \wedge e^b$ and inserting
Eqn(\ref{AlanDenk3}) into that equation, we obtain
 \ba
     \D \Sigma_a{}^b - \tau_a \wedge e^b +M \delta^b_a *1 =0
 \ea
where $e^a \wedge *e^b = \eta^{ab} *1$ is used. We lower the index $b$
by paying special attention in order to be able to use symmetry
arguments,
 \ba
    \kappa \D *Q_{ab} + 2 \kappa Q^c{}_b \wedge *Q_{ac} - \tau_a
    \wedge e_b + M \eta_{ab} *1 =0 \label{AlanDenk4}
 \ea
where we write $\Sigma_{ab}$ in terms of $Q_{ab}$ through
(\ref{Sigma_a^b}). It is clear that $Q^c{}_b \wedge *Q_{ac} =
Q^c{}_a \wedge *Q_{bc}$. Besides, it may be seen by using
Eqn(\ref{tauA}) that $\tau_a  \wedge e_b = \tau_b \wedge e_a $.
Thus, Eqn(\ref{AlanDenk4}) is definitely symmetric. As
the off-diagonal elements give
 \ba
   \kappa \d *Q_{ab}  - \kappa (\imath_aQ^{cd}) (\imath_bQ_{cd}) *1
   =0 \; , \quad (a\neq b), \label{alandenk0}
 \ea
the trace of Eqn(\ref{AlanDenk4}) yields
  \ba
   \kappa \d*Q  + \kappa  Q^{cd} \wedge *Q_{cd} + 4M *1 =0
   \label{alandenkiz}
 \ea
where we used $\Lambda^a{}_b = Q^a{}_b$ in the expression
$\D*Q_{ab}$. Finally, the covariant exterior derivative of
Eqn(\ref{AlanDenk2}) produces
 \ba
   \D\tau_a - MQ \wedge *e_a =0 \label{AlanDenk5}
 \ea
where we use $\D*e_a = -Q \wedge *e_a$  via Eqn.(\ref{ident:1})
and $\D^2\lambda_a = -R^b{}_a \wedge \lambda_b =0$. As a
conclusion, the last three equations are our field equations derived from the locally gauge invariant STPG Lagrangian (\ref{LagranjYerel2}).

 \subsection{A Conformally Flat Solution}

Let us follow the steps in Section \ref{kisimSTPG} as an
application. We firstly make a conformally flat metric ansatz in the Cartesian coordinate chart, $x^\alpha = (t,x,y,z)$,
 \ba
   g=e^{2\psi} (-\d t^2 + \d x^2 + \d y^2 + \d z^2)
 \ea
where $\psi=\psi(t,x,y,z)$ is the unknown function to determined by the field equations. Thus, $ g_{\alpha \beta} =
e^{2\psi} \delta^a_\alpha \delta^b_\beta \eta_{ab} $. In the
second step, we choose the co-frame and the connection 1-forms as
$e^\alpha = \d x^\alpha$ and $\Lambda^\alpha{}_\beta =0 $,
respectively,  in the coordinate gauge. Now definitely $
R^\alpha{}_\beta =0 $, $ T^\alpha =0 $, $ Q_{\alpha \beta} = -
\frac{1}{2} \d g_{\alpha \beta} = -\d \psi g_{\alpha \beta}$. In
the third step, we write the orthonormal co-frame $e^a = e^\psi
\delta^a_\alpha \d x^\alpha$ such that the vierbein are $h^a{}_\alpha
= e^\psi \delta^a_\alpha$ and the inverses are $h^\alpha{}_a =
e^{-\psi} \delta^\alpha_a$. In the fourth step
 \ba
  \Lambda^a{}_b = h^a{}_\alpha \d h^\alpha{}_b = -\d \psi \delta^a_b
  \, , \; \;  R^a{}_b = 0 \, , \;\; T^a = 0 \, , \;\; Q^a{}_b = -\d \psi
  \delta^a_b \, . \label{Qab}
 \ea
Then we will insert the nonmetricity given by
Eqn(\ref{Qab}) into the field equations
(\ref{alandenk0})-(\ref{AlanDenk5}) and then we will solve them
for $\psi$.

When we put the nonmetricity (\ref{Qab}) into the symmetric
field equation (\ref{alandenk0}) we obtain
 \ba
     (\partial_\alpha \psi) (\partial_\beta \psi) =0
 \ea
where it is used that $g_{\alpha \beta}=0$ for $\alpha \neq
\beta$. We also insert the nonmetricity (\ref{Qab}) into the trace
field equation (\ref{alandenkiz}). Thus we obtain
 \ba
     \partial^\alpha \partial_\alpha \psi +  (\partial^\alpha \psi) (\partial_\alpha \psi) - M/\kappa
     =0 \; .
 \ea
Finally we substitute Eqn(\ref{Qab}) into Eqn(\ref{AlanDenk5}) to
get
 \ba
   4 (\partial_\alpha \psi) [\partial^\beta \partial_\beta \psi +  (\partial^\beta \psi) (\partial_\beta \psi) - M/\kappa]
   =0
 \ea
where $\partial^\alpha \partial_\alpha = g^{\alpha \beta}
\partial_\alpha \partial_\beta$ together with $\partial_\alpha := (\partial / \partial t , \partial /
\partial x , \partial / \partial y , \partial / \partial z )$.
Thus one has to solve the following equations
 \ba
 (\partial_\alpha \psi) (\partial_\beta \psi) &=&0 \; , \quad \alpha \neq \beta, \label{denk0} \\
  -\ddot{\psi} + \nabla^2 \psi -{\dot{\psi}}^2 + (\vec{\nabla}\psi).(\vec{\nabla}\psi) - (M/\kappa) e^{2\psi} &=& 0 \label{denk1konf}
 \ea
where dot and $\vec{\nabla}$ stand for $t$-derivative and spatial
gradient operator, respectively. According to the first equation
$\psi$ can be dependent only on one coordinate. For instance, let
$\psi=\psi(t)$ where $t$ is the time coordinate. Now
Eqn(\ref{denk1konf}) takes the form
          \ba
            \ddot{\psi} +  \dot{\psi}^2 + (M/\kappa) e^{2\psi} =0
            \, . \label{eq:psit}
          \ea
After $\psi(t) =  \ln y(t)$, the equation turns out to be
          \ba
           \ddot{y} + (M/\kappa)y^3=0 \, .
          \ea
Thus we obtain $t= \int \left( b^2 -(M/2\kappa) y^4 \right)^{-1/2}
\d y$ where $b$ is a constant. If $b\neq 0 $, this integral can be
written in terms of the elliptical integrals. Setting $b=0$, an exact
solution is obtained as  $y(t)= -(a+ \sqrt{-M/2\kappa}t)^{-1}$
where $a$ is a constant. Now if one more coordinate transformation
is performed as $T=\int e^{\psi(t)} \d t = \int y(t) \d t$, then
the result becomes $ (a+ \sqrt{-M/2\kappa}t)^{-1} = c
e^{\sqrt{-M/2\kappa}T}$. With a suitable choice of
the constant $c$, the metric is expressed as
           \ba
              g= -\d T^2 + S^2(T) (\d x^2 + \d y^2 + \d z^2)
           \ea
where $S^2(T)=e^{2\psi[t(T)]}=  e^{\sqrt{-2M/\kappa}T}$. Here
$\sqrt{-2M/\kappa}$ is  a real constant when the signs of
constants are chosen  appropriately.

 \section{The Result}

Although the theory of general relativity is mathematically elegant,
we have persuasive motivations caused by recent astrophysical and
cosmological observations in order to modify it. One way of achieving that is to go beyond the Riemannian geometry.
In this context we study the theory of STPG developed by only the nonmetricity tensor.
Even though the physical aspects of the dynamical nonmetricity are not
easy to interpret, it can be important for understanding of some
astrophysical events like dark matter \cite{tucker1998}.
Besides dark matter, astrophysical applications of nonmetricity, indeed specifically in the STPG, have been proposed recently.
Exact solutions that describe inflation with nonmetric UV corrections and dark energy with nonmetric IR corrections were presented in \cite{koivisto2017}.
Therefore, these kinds of theoretical studies are needed to be
carried out. Here we built a locally gauge invariant
STPG model in the symmetric teleparallel geometry, which has nonzero nonmetricity,
but neither curvature nor torsion. The subject is treated from the perspective of the gauge
theory. In similar work \cite{koivisto2018} the author proposes a gauge interpretation to STPG from a physical point of view:
Since the equivalence principle allows eliminating locally gravitation, it must be an integrable gauge theory.
In the current approach, the metric represents the gravitational field and
the connection corresponds to the gauge potential. Thus, the locally gauge invariant
Lagrangian containing kinetic and mass terms is written and
the variational field equations are derived. While a solution is
sought to these equations, the metric and the connection ansatzes are made
independently in the literature. In this work we show that
only a metric ansatz is enough to begin to look for a non-trivial
solution by choosing the full connection zero in the coordinate frame. As a concrete application, a set of conformally flat solutions to
the field equations is obtained explicitly.
In the work \cite{Hohmann2018} the author determines the possible wave polarizations in STPG by linearizing the field equations.
Accordingly, one possible extension of this work is to seek exact gravitational wave solutions to more general STPG models in our future researches.

 \section*{Acknowledgement}

This work was supported by the Scientific Research Coordination Unit of Pamukkale University under the project number 2018HZDP036. The author thanks to Murat Sar{\i} and \"Ozcan Sert for the fruitful discussions, and the anonymous referees for their guiding criticisms.

\end{document}